%% file: IAU237_IRL.tex
\documentclass{iau}
\usepackage{graphicx}
\usepackage{natbib}
\usepackage{tabularx}
\usepackage{amsmath}

\graphicspath{{./fig/}{./png/}}

\input{macros.tex}

\makeatletter
\input{aas_macros.tex}

\makeatother

\title[Formation of sunspots] 
{A new look at sunspot formation using theory and observations}

\author[Illa R. Losada]   
{Illa R. Losada$^{1,2,3,4}$,
 J\"orn Warnecke$^5$,
 Kolja Glogowski$^4$,
\\
 Markus Roth$^4$,
 Axel Brandenburg$^{1,2,6}$,
 Nathan Kleeorin$^{7,1}$,
\\
 \and
 Igor Rogachevskii$^{7,1}$
 }
\affiliation{$^1$Nordita, KTH Royal Institute of Technology and Stockholm University,
Roslagstullsbacken 23, SE-10691 Stockholm, Sweden, email: {\tt
  illa.rivero.losada@gmail.com} \\
$^2$Department of Astronomy, Stockholm University, SE-10691 Stockholm,
Sweden\\
$^3$Nordic Optical Telescope, Apartado 474, E-38700 Santa Cruz de La Palma, Spain\\
$^4$Kiepenheuer-Institut f\"ur Sonnenphysik, Sch\"oneckstra{\ss}e 6, D-79104,
Freiburg, Germany\\
$^5$Max Planck Institute for Solar System Research,
Justus-von-Liebig-Weg 3, D-37077 G\"ottingen, Germany\\
$^6$Laboratory for Atmospheric and Space Physics,
JILA and Department of Astrophysical and Planetary Sciences,
University of Colorado, Boulder, CO 80303, USA\\
$^7$Department of Mechanical Engineering, Ben-Gurion University of the
Negev, POB 653, Beer-Sheva 84105, Israel\\
}

\pubyear{2016}
\volume{327}  
\jname{IAUS 327: Fine Structure and Dynamics of the Solar Atmosphere}
\editors{A.C. Editor, B.D. Editor \& C.E. Editor, eds.}
\begin{document}

\maketitle

\begin{abstract}
Sunspots are of basic interest in the study of the Sun.
Their relevance ranges from them being an activity indicator of magnetic fields
to being the place where coronal mass ejections and flares erupt.
They are therefore also an important ingredient of space weather.
Their formation, however, is still an unresolved problem in solar
physics. Observations utilize just 2D surface information near the spot,
but it is debatable how to infer
deep structures and properties from local helioseismology.
For a long time, it was believed that flux tubes rising from the
bottom of the convection zone are the origin of the bipolar sunspot
structure seen on the solar surface.
However, this theory has been challenged, in particular recently by new
surface observation, helioseismic inversions, and numerical models of
convective dynamos.
In this article we discuss another theoretical approach to the
formation of sunspots: the negative effective magnetic pressure
instability.
This is a large-scale instability, in which the total (kinetic plus
magnetic)
turbulent pressure can
be suppressed in the presence of a weak large-scale magnetic field,
leading to a converging downflow, which eventually concentrates
the magnetic field within it.
Numerical simulations of forced stratified turbulence have been able to
produce strong super-equipartition flux concentrations, similar to
sunspots at the solar surface.
In this framework, sunspots would only form close to the surface
due to the instability constraints on stratification and rotation.
Additionally, we present some ideas
from local helioseismology,
where we plan to use the Hankel analysis to study
the pre-emergence phase of a sunspot and to
constrain its deep structure and formation mechanism.

\end{abstract}

\firstsection 
\section{Introduction}

Over the past four centuries, the study of sunspots has advanced significantly
from simple counts and drawings to detailed monitoring from space.
A lot of effort has gone into the compilation of a coherent sunspot
record \citep{Clette14}.
Despite the long-term record and the high spatial and temporal
resolution nowadays available, the theoretical understanding
of the processes leading to the formation of sunspots is still limited.

One of the first theoretical descriptions of sunspot formation and
evolution was provided by \cite{Par55b}.
He argues that the solar dynamo produces magnetic fields in
the form of concentrated flux tubes, which are able to rise through
the convection zone to the surface by magnetic buoyancy.
An initially horizontal flux tube is expected to
break through the surface, forming a typical bipolar sunspot pair.
Originally, \cite{Par55b} expected the flux tubes to be just $20\Mm$
below the surface, but later he argued that they should be
near the overshoot layer \citep{Parker75}.

There has been a range of different approaches to this scenario: from thin flux
tube models \citep[e.g.][]{caligari95} to detailed simulations
of the emergence of horizontal field over the last few megameters
\citep[e.g.][]{rempel2014}.
\cite{caligari95} studied the process of the emergence from the tachocline,
concluding that a critical magnetic field strength of
$\approx 10^5\G$ in the subadiabatic layer is needed for the
magnetic buoyancy to kick in and raise the tube.
They also studied some
of the observed properties of active regions, such as the tilt
angle and the asymmetry of the inclination angle and the velocity,
as well as the latitudinal emergence.
On the other hand, \cite{rempel2014} studied in detail the rise of
a pre-injected magnetic flux tube near the surface and the
characteristics of bipolar spot formation in the photosphere.
Turbulent flows tend to fragment and disperse the
injected coherent magnetic flux rope, leading to the decay of the
spot.
Further examples of theoretical models of flux tube emergence can be found
in \cite{Cheung+Isobe:14}.

\section{Change of paradigm}

This traditional and most studied paradigm to explain the formation of
sunspots is part of a more general picture of the Sun.
In this framework, which is also referred to as the Babcock-Leighton dynamo
\citep{Babcock61,Leighton64,Leighton69}, the poloidal magnetic
field gets sheared into a toroidal one by the differential rotation of the Sun.
This toroidal field resides at the tachocline.
There, the layer can be stable enough to allow for a magnetic field of
$\sim 10^5\G$, which it needs to survive its rise as a coherent tube through the
convection zone and to break through the surface, forming spots.
Therefore, these flux ropes of magnetic field are global features that
feed the dynamo of the Sun, and the sunspots are deeply rooted within
the convection zone.
One of the problems of the stability of these tubes is convection itself.
The turbulence tends to diffuse and destroy the tubes, so they
have to be strong enough to survive and rise throughout the entire
convective envelope.

We could also imagine a different framework, where the dynamo may still be
of Babcock-Leighton type, but it operates in the bulk of the convection
zone and causes most of the global features of the Sun.
Sunspots form under certain favorable conditions (dynamo strength,
stratification, scale separation of the convective cells, etc)
as the result of an instability.
On top of this, we could also picture a scenario where the turbulence
itself is driving a large-scale instability and forms the magnetic field
concentrations seen as sunspots \citep{KRR89,KRR90,KMR96}.
We argue that the negative effective magnetic pressure instability (NEMPI)
is the instability operating at this level; see \cite{BRK16} for a recent review.
These two scenarios are summarized in \Tab{tab:paradigm}.

\begin{table}
 \begin{center}
  \caption{Main characteristics of the presented paradigms on the formation of
    sunspots}
  \label{tab:paradigm}
  \begin{tabularx}{\columnwidth}{ X | X | X}
\hline
                & Traditional approach & Alternative approach  \\
\hline
Dynamo type         & tachocline flux-transport & distributed \\
Role of sunspots & part of dynamo & natural outcome\\
Origin of sunspots & rising flux tubes & turbulence \& stratification \\
Sunspot formation depth  &  tachocline       & near the surface  \\
Turbulent convection      & sunspot decay        & sunspot formation
\\
\hline
\end{tabularx}
 \end{center}

\end{table}

\section{Negative effective magnetic pressure instability (NEMPI)}

\subsection{MHD simulations}

In order to introduce the mechanism responsible for the formation of
magnetic field concentrations, we have to go back to the equations used
to study the dynamics of the plasma:
the magnetohydrodynamics (MHD) equations.
We then use simulations to solve the relevant set of equations.
One of the main problems in numerical simulations of the Sun
is the inability to use realistic parameters.
The solar  Reynolds number, $\Rey$, related to the amount of turbulence in the system,
can span from $10^{10}$ to $10^{15}$, requiring a numerical resolution not yet
achievable in super-computers.
We, then, have to use a much smaller value of Reynolds number
and study how the system scales with increasing values of $\Rey$.

Here, we solve the MHD equations using both direct numerical
simulations (DNS) and mean-field simulations (MFS).
The DNS scheme solves the full set of equations included in the problem,
for a given numerical resolution.
A drawback of this approach is that
it can be difficult to capture the effects of small-scale fluctuations.
This is different using MFS,
where the problem is solved for the large-scale quantities and the
contribution of small-scale quantities are parameterized in terms of
large-scale quantities.
This allows us to make assumptions for the derivation of the parameterizations and
consider only the relevant terms.
Therefore, comparing DNS results with known MFS physics will lead to a
better understanding of the physical system.

\subsection{DNS}

The full set of equations we solve within a DNS scheme is:

\begin{itemize}
 \item Continuity equation: $\partial\rho/\partial t=-\nab\cdot(\rho\UU)$,
 \item Momentum equation: $\DD\UU/\DD t=-{1\over\rho}\nab P_{\rm gas}+{1\over\rho}\JJ\times\BB
+\ff+\grav+\FF_\nu$,
 \item Induction equation: $\partial \BB/\partial t = \nab \times (\UU \times \BB
 - \eta \mu_0\JJ)$,
\end{itemize}
where:
$\DD/\DD t \equiv \partial/\partial t+\UU\cdot\nab$ is the advective derivative,
$P_{\rm gas}$ is the gas pressure,
$\nu$ is the kinematic viscosity, $\eta$ is the magnetic diffusivity,
$\BB=\BB_0+\nab\times\AAA$ is the magnetic field,
$\BB_0=(0,B_0,0)$ is a weak imposed uniform field,
$\JJ=\nab\times\BB/\mu_0$ is the current density,
$\mu_0$ is the vacuum permeability,
$\FF_\nu=\rho^{-1}\nab\cdot(2\nu\rho\SSSS)$ is the viscous force,
${\sf S}_{ij} =\half(\partial_j U_i+\partial_i U_j)-\onethird\delta_{ij}\nab\cdot\UU$ is the traceless rate-of-strain tensor,
$\ff$  is the forcing function (random, white-in-time, plane,
nonpolarized waves with a certain average wavenumber $\kf$),
$\grav=(0,0,-g)$ is the gravitational acceleration.

We measure distance in terms of the density scale height, $H_\rho=\cs^2/g$,
magnetic field in terms of the equipartition field strength,
$\Beqz=\sqrt{\mu_0\rho_0} \, \urms$,
and time in terms of the turbulent-diffusive time, $t_\eta = (\etat k_1^2)^{-1}$
(which is proportional to the
eddy turnover time, $\tau_{t0}=(\urms \kf)^{-1}$, since the turbulent
magnetic diffusivity is assumed to the equal to the turbulent
viscosity: $\etat= \nut=\urms/3\kf$).

\cite{BKKMR11} solved these equations in a small 3D box, with a
scale separation of $\kf/k_1 = 15$, strong stratification with a density
contrast of $\sim 535$,
an initial magnetic field along the
horizontal $y$-axis of $B_0/\Beqz =0.05$, a fluid Reynolds number
$\Rey\equiv\urms/\nu\kf$ of 36,
a magnetic Prandtl number $\Pm=\nu/\eta$ of 0.5,
a magnetic Reynolds number, $\Rm=\Pm\Rey=36$,
and detected, for the first time, the spontaneous concentration of magnetic field.
This concentration is a consequence of the large-scale instability, i.e.,
NEMPI, which is described in terms of mean-field equations.

\subsection{MFS}

Mean-field equations emerge as a result of the decomposition of all
quantities as averages (means) and fluctuations, $\WW = \meanWW + \ww$.
Applied to the MHD equations, this decomposition yields:

\begin{itemize}

 \item Continuity equation: $\partial\meanrho/\partial t=-\nab\cdot(\meanrho\meanUU)$,
 \item Momentum equation: $\partial\meanUU/\partial t+\meanUU\cdot\nab\meanUU=
-{1\over\meanrho}\nab\meanp +\grav+\meanFFFF_{\rm M}+\meanFFFF_{\rm K}$,
 \item Induction equation: $\partial \meanBB/\partial t = \nab \times
(\meanUU \times \meanBB + \overline{\mathbf{u}\times \mathbf{b}})  +\eta \nab^2
\meanBB$,
\end{itemize}
where overbars denotes averages, $\meanFFFF_{\rm M}$ is the mean Lorentz
force and $\meanFFFF_{\rm K}$ is the total (turbulent plus microscopic)
viscous force, and the turbulent viscous force is determined by Reynolds stresses.

\subsubsection{From induction equation: dynamo instability}

As a well-known example of mean-field theory,
the mean electromotive force $\overline{\uu\times \bb}$
in the mean field induction equation is caused by turbulent effects
which can result in the generation of a large-scale magnetic field \citep{SKR66,Mof78,KR80}.
In the case of isotropic and homogeneous turbulence, the effect of
fluctuations can be parameterized as
\EQ
\overline{\uu\times \bb}=\alpha \meanBB -\eta_t\meanJJ.
\EN
Therefore, if $\alpha \neq 0$, there can be a generation of
a large-scale magnetic field.
This effect is thought to be responsible for generating the Sun's
large-scale magnetic field \citep{Stix1976,Dikpati2001,BS05}.

\subsubsection{From momentum equation: NEMPI}

For the current discussion, we are not interested in explaining the generation of a
magnetic field, but in the study of a concentration of an existing
small large-scale magnetic field into the form of surface spots.
Therefore, we use the momentum equation:
\begin{equation}
 {\partial\meanUU\over\partial t}+\meanUU\cdot\nab\meanUU=
-{1\over\meanrho}\nab\meanP_{\rm gas}+\grav+\meanFFFF_{\rm M}+\meanFFFF_{\rm K}.
\end{equation}
The total pressure $P_{\rm Tot}$ in the system
is the gas pressure, $P_{\rm gas}$,
plus the mean magnetic pressure, $\meanBB^2 / 2\mu_0$,
and the total (kinetic plus magnetic) turbulent pressure,
$p_{\rm turb}=p_{\rm turb}^{(0)} + p_{\rm turb}^{(B)}$, where
the total turbulent pressure has a contribution, $p_{\rm turb}^{(B)}$,
which depends on the mean magnetic field and
one, $p_{\rm turb}^{(0)}$, which is independent on $\meanBB$.
We define the effective magnetic pressure $\Peff=\meanBB^2 / 2\mu_0 + p_{\rm turb}^{(B)}$,
so that the total pressure is:
\begin{equation}
P_{\rm Tot} = P_{\rm gas} + {\meanBB^2 \over 2\mu_0} + p_{\rm turb} \equiv \tilde P_{\rm gas}
+ \Peff ,
\label{eq:PT}
\end{equation}
where $\tilde P_{\rm gas}=P_{\rm gas} + p_{\rm turb}^{(0)}$.
The turbulent pressure for isotropic turbulence can be written as:
\EQ
p_{\rm turb} = {E_{\rm M}\over 3} + {2 E_{\rm K}\over 3}
= {2\over 3} (E_{\rm K}+ E_{\rm M}) - {1\over 3} E_{\rm M}.
\EN

Therefore, if the total energy $E_{\rm K}+E_{\rm M}$ is conserved,  an increase
in the turbulent magnetic energy, $\delta E_{\rm M} > 0$, suppresses the turbulent pressure,
$\delta p_{\rm turb} = - (1/3) \delta E_{\rm M} < 0$,
resulting in a negative contribution to the total pressure.
For strongly anisotropic turbulence, $\delta p_{\rm turb} = - \delta E_{\rm M} < 0$,
and the suppression of the turbulent pressure is stronger \citep{RK07,BRK16}.
The contribution to the total turbulent pressure, $p_{\rm turb}^{(B)}$,
is parameterized as $p_{\rm turb}^{(B)} = -\qp(\meanBB) \, \meanBB^2 / 2\mu_0$,
so that the effective magnetic pressure is \citep{KRR89,KRR90,KMR93,KR94,RK07}:
\EQ
\Peff = \left[1-\qp(\meanBB)\right]{\meanBB^2 \over 2\mu_0}.
\label{eq:peff}
\EN
Substituting the effective magnetic pressure of \Eq{eq:peff} into \Eq{eq:PT},
we obtain the following expression for the total pressure:
\begin{equation}
P_{\rm Tot} =\tilde P_{\rm gas} + \left(1-\qp(\meanBB)\right){\meanBB^2 \over 2\mu_0}.
\label{eq:PT2}
\end{equation}
In cases where the function $\qp > 1$,
the contribution of the effective magnetic pressure to the total pressure
is negative,
and the gas pressure should increase if the total
pressure is balanced sufficiently rapidly with its surroundings.
Indeed, a small increase in the magnetic field in the upper layer
results in the increase of the gas pressure in this region
Therefore, there is a positive gas pressure difference between
the upper and lower layer, which drives the downflows.
This eventually concentrates the vertical magnetic field
within it at the expense of turbulent energy, and the NEMPI is excited.

\paragraph{\it Effective magnetic pressure computation and parameterization:}

The effective magnetic pressure relates the mean magnetic energy to the
turbulent one through the function $\qp$.
We can compute the ad-hoc form of this function by comparing a simulation
with an imposed mean magnetic field to a simulation without this field,
and extract the effect of the mean magnetic field
in the effective magnetic pressure.

\cite{Kemel12a} fitted the DNS results,
and approximated the function $\qp(\beta)$ by:
\EQ
\qp(\beta)={\qpz\over1+\beta^2/\betap^2}
={\betastar^2\over\betap^2+\beta^2},
\label{qp-apr}
\EN
where $\beta = \meanBB^2/\Beq^2$, and $\qpz$, $\betap$,
and $\betastar=\betap\qpz^{1/2}$ are constants.
Also, \cite{Kemel13a} established an approximation
for the growth rate of the instability, in the case of an isothermal
atmosphere and $\betastar \gg \beta \gg \betap$:
\EQ
{\lambda \over \etat k^2} \approx 3 \betastar {\kf/k \over kH_\rho} -1 .
\EN
Studying the large-scale instability, we can infer the importance
of stratification and the scale separation at the onset of the instability.
Therefore, we need three main conditions
to excite the large-scale instability:
\begin{itemize}
 \item The gradient of effective magnetic pressure must be negative.
 \item Strong enough stratification (small density scale height) $kH_\rho$.
 \item Large enough scale separation in the system $\kf/k$.
\end{itemize}
These conditions have already been
demonstrated in the DNS of \cite{BKKR12,BKR13}
and their results set up a starting point for further
numerical experiments towards solar-like simulations.

\subsection{Some results}

To go a step further towards a solar-like model, we focus on two
aspects that influence the development of the instability present in
the Sun: rotation and a basic representation of a corona envelope.

\begin{figure}[h]
\begin{center}
 \includegraphics[width=\columnwidth]{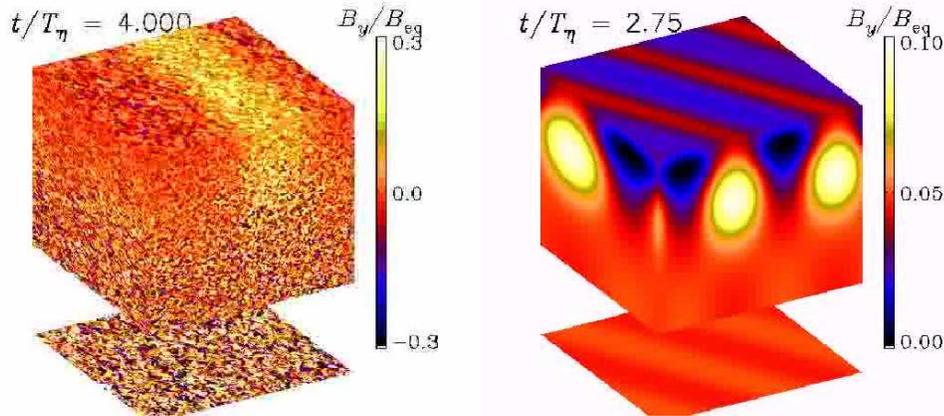}
 \caption{$B_y$ at the periphery of the computational domain for DNS
   (\emph{left}) and MFS (\emph{right}) with $\Co = 0.03$. Figures adapted from
   \cite{LBKMR2012,LBKR2013a}.}
   \label{fig:DNSMFS}
\end{center}
\end{figure}

\subsubsection{Effects of rotation}

We study the effects of rotation by adding the angular velocity term, $-2\OO\times\UU$ in the momentum equation,
such as $\OO=\Omega\left(-\sin\theta, 0, \cos\theta\right)$,
where $\Omega$ is the rotation rate and
$\theta$ is the colatitude of the Sun ($\theta=0$ at the poles).
Therefore, we use the following equations of motion in DNS and MFS:
\EQA
\text{DNS:} \quad {\DD\UU\over\DD t}&=&-2\OO\times\UU -\cs^2\nab\ln\rho
+{1\over\rho}\JJ\times\BB+\ff+\grav
+\FF_\nu . \\
\text{MFS:} \quad {\DD\meanUU\over\DD t}&=&-2\OO\times\meanUU
-\cs^2\nab\ln\meanrho+{1\over\meanrho}\meanJJ\times\meanBB+\nab(\qp\meanBB^2/2\mu_0)+
\grav
+\meanFF_{\nut}.
\ENA
As explained in \cite{LBKMR2012, LBKR2013a},
we set up MFS and DNS of boxes of size $L^3$, and impose a weak
horizontal uniform magnetic field in the $y$ direction,
$\BB_0$.
The computational domain has periodic boundary conditions in the $x$ and $y$ directions,
while we impose stress-free perfect conductor boundary conditions in
the $z$ directions.

We solve these equations in a small 3D box, with a scale separation of
$\kf/k_1$, a strong stratification of a density contrast of $\sim 535$, a initial
magnetic field along the y-axes of $B_0/\Beq = 0.05$,
a fluid Reynolds number $\Rey\equiv\urms/\nu\kf$ of 36,
a magnetic Prandtl number $\Pm=\nu/\eta$ of 0.5,
and a magnetic Reynolds number, $\Rm=\Pm\Rey=18$.

\begin{figure}[ht]
\begin{center}
 \includegraphics[width=0.7\columnwidth]{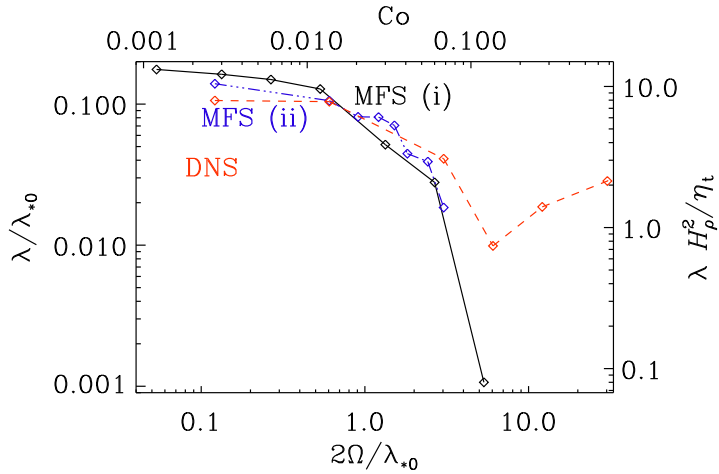}
 \caption{Dependence of the growth rate of the instability on the rotation rate,
for DNS (red dashed line) and MFS with (i): $\qpz=20$ and
$\betap=0.167$ (black solid line), and (ii): $\qpz=32$ and $\betap=0.058$
(blue dash-dotted line). Figure adapted from \cite{LBKR2013a}.}
   \label{fig:omdep}
\end{center}
\end{figure}

\begin{figure}[ht]
\begin{center}
 \includegraphics[width=0.4\columnwidth]{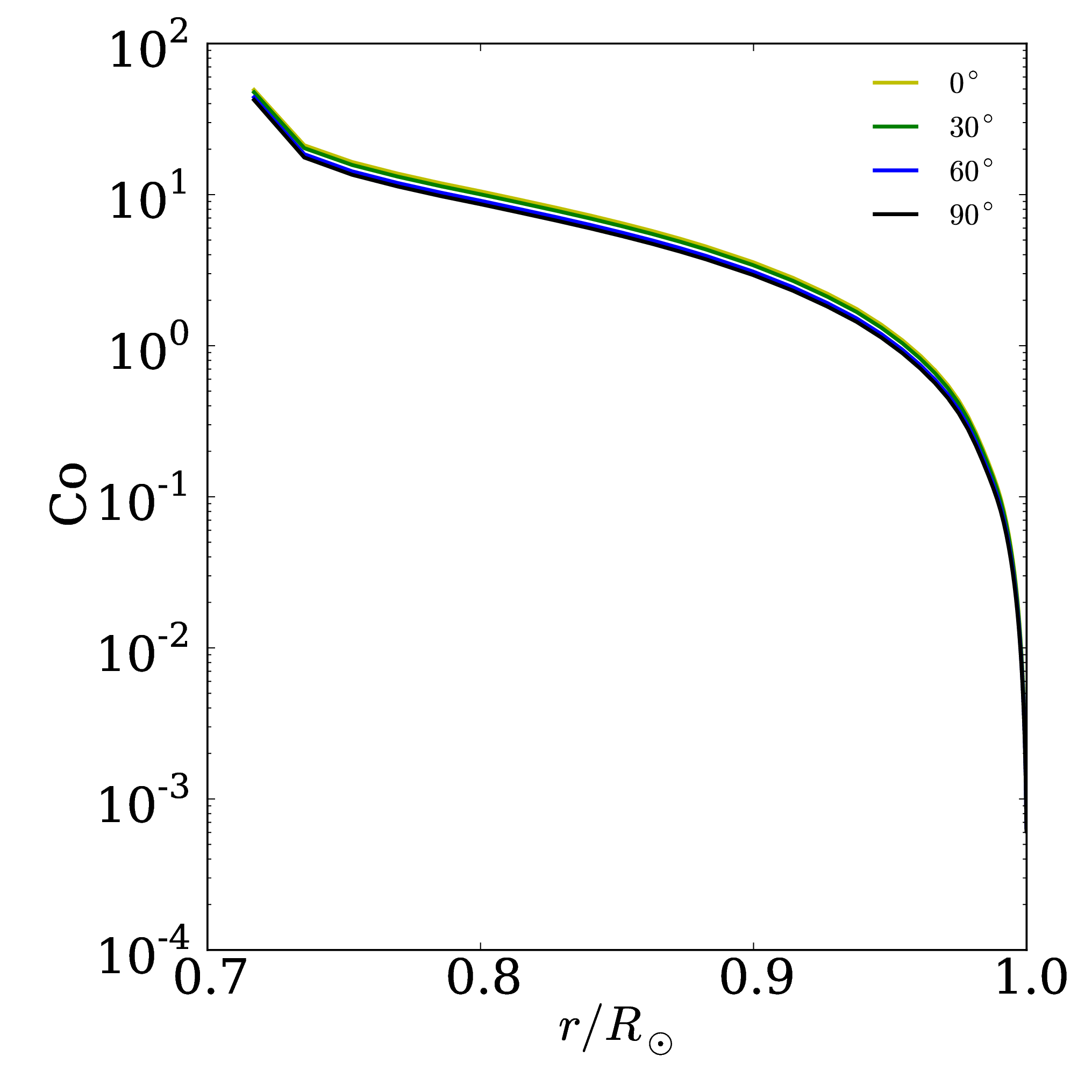}
 \caption{Coriolis number profile in the convection zone, computed form the mixing
 length model of \cite{Spruit74}.}
   \label{fig:co}
\end{center}
\end{figure}

In the presence of rotation, we observe the formation of magnetic
field concentrations, as seen in \Fig{fig:DNSMFS}.
We quantify the growth rate of the large-scale instability,
normalized by the quantity $\lambda_{\ast0}\equiv\betastar\urms/H_\rho$.
In the DNS we can directly compute the value of $\betastar$, whereas in the MFS
it is an input parameter of the simulation.
We study two different sets of values in MFS:
(i) with $\qpz=20$ and $\betap=0.167$, motivated by the results of \cite{Kemel13a},
and (ii) with $\qpz=32$ and $\betap=0.058$, motivated by the DNS fits.
As one of the results, \Fig{fig:omdep} shows the decrease of the
growth rate of the instability when we increase
the Coriolis number in all these three cases.
In a solar context, however, where the Coriolis number
strongly decreases towards the surface
(see \Fig{fig:co}), the limitation in Coriolis number of the
instability constrains the area where the instability
can operate close to the surface,
where the Coriolis number is small enough.
As another result, we can see a disagreement between DNS and MFS
results for values of the Coriolis number larger than $\approx 0.1$.
Such values of the rotation rate enable the production of kinetic
helicity in the system, and therefore an $\alpha$-effect.
We see the development of a Beltrami-like magnetic field generated by
the large-scale dynamo together with the NEMPI instability.
\cite{Jabbari2014} demonstrated that an $\alpha^2$ mean-field
dynamo is responsible for the onset of dynamo and the presence of the instability
for even larger values of the Coriolis number.
In the presence of rotation
(angular velocity above a certain threshold) the system is
unstable with respect to two different instabilities: the large-scale dynamo
instability and NEMPI.
In general, we expect these instabilities to couple and therefore make
the calculation of growth rate much more challenging.

\subsubsection{Two-layer stratified model}

\begin{figure}[ht]
\begin{center}
 \includegraphics[width=\columnwidth]{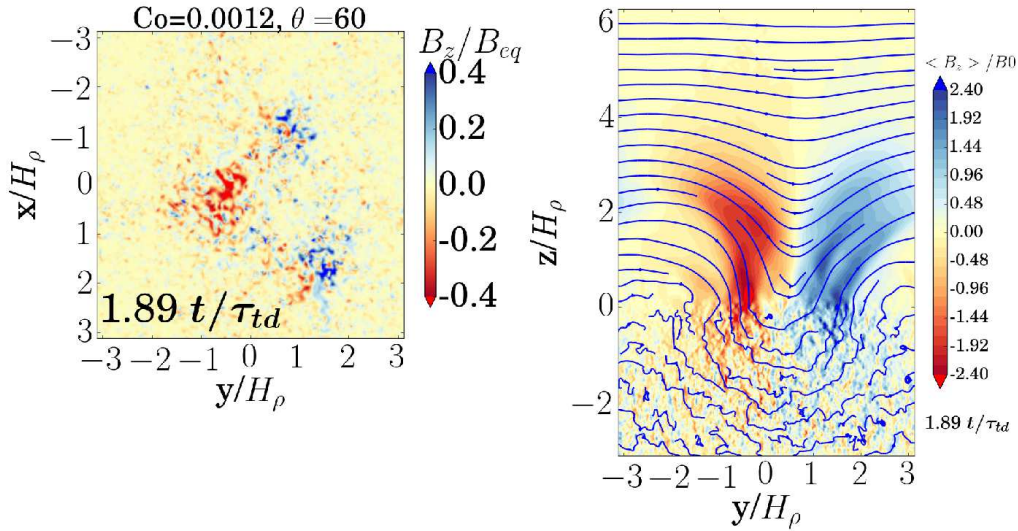}
 \caption{Structure formation in a two-layer model. We show the $z$
 component of the magnetic field at the surface, normalized by the
 surface equipartition field strength $\Beq$ (\textit{left}),
 and in the $zy$ plane, averaged over the $x$ direction and
 normalized by the imposed field $B_0$ (\textit{right}).
 The magnetic field lines are plotted in blue.
 Resolution is $192^2 \times 384$ grid points.}
   \label{fig:bb3}
\end{center}
\end{figure}

Our previous models used
either vertical field or perfect conductor condition
at the top boundary for the magnetic field, when we either impose a
vertical or horizontal magnetic field, respectively.
These constrain the dynamics of NEMPI.
One way to avoid this problem and to reproduce solar-like conditions
is to include a non-turbulent region on top of turbulent layer.
We call this layer a coronal envelope as it mimic a idealized corona
on top of the convection zone of the Sun.
Because of the coronal envelope the magnetic field evolves and changes
freely at the surface (boundary between turbulent and non-turbulent
layer).
This approach have been used successfully for dynamo simulation of forced turbulence
\citep{WB2010,WBM2011} and turbulent convection \cite{WKMB12,WKMB13,WKKB16}.
The first attempt to combine this two-layer with the study of NEMPI
can be found in \cite{WLBKR2013, WLBKR2015}.

Now the initially uniform $y$ component of magnetic field
is able to change orientation at the surface,
resulting in the spontaneous formation
of bipolar regions, which form, evolve and disappear at the surface.
We study the effects of slow rotation in a two-layer stratified model,
using different resolutions, rotation rates and colatitudes.
\Fig{fig:bb3} shows the formation of such bipolar regions in the case of slow
rotation at the surface (left panel), and a vertical average (right panel)
at the time of the maximum.
Now, with the corona envelope, the instability is able to change the
orientation of the initial  homogeneous $y$ directed imposed field,
generating a field in the vertical direction that concentrates in the
form of bipolar regions.

\begin{figure}[ht]
\begin{center}
 \includegraphics[width=0.9\columnwidth]{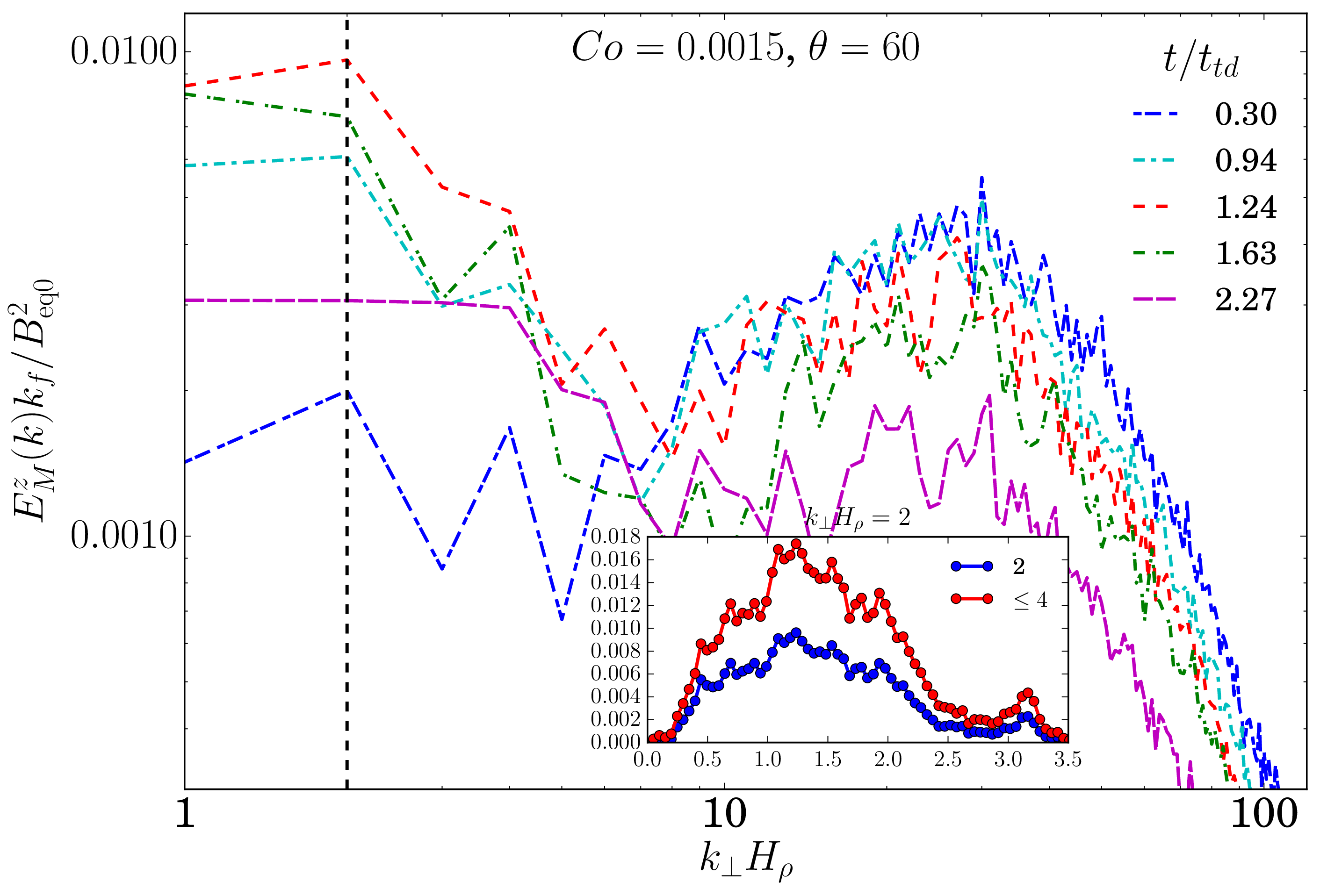}
 \caption{Power spectra of the $z$ component of the
   magnetic field, $E^z_{\rm M}$
 (normalized by the equipartition field strength at the surface $\Beqz$),
 over different scales of the system ($k_\perp H_\rho$)
 and time evolution of the energy at a wavenumber $k_\perp = 2$.
 Same simulation as in \Fig{fig:bb3}}.
   \label{fig:em}
\end{center}
\end{figure}

A good proxy to understand how the instability operates is the
evolution of the magnetic energy spectrum shown in \Fig{fig:em}.
Initially, the magnetic energy is concentrated at the forcing scale $\kf/k_1 = 30$.
When the bipolar region starts forming, the energy is transferred to a
larger scale, with a maximum at $k=2$, when we see the bipolar
structure appear at the surface.
The instability is then suppressed, and the structures decay,
transferring back the energy to smaller scales in the system.

As in previous results, we also see the suppression of the instability
as we increase the Coriolis number, but now the presence of the corona
layer helps the instability survive even at higher rotation rates.

\section{Local helioseismology}

Helioseismology is the only technique that allows us to study the solar interior
and infer its inner structure and flows.
Furthermore, local helioseismology allows the study of small patches of
the Sun,
opening the possibility to compute flows and  magnetic fields.
Local helioseismology make use of different methods, including
ring-diagram
analysis, time-distance helioseismology, helioseismic holography, and the
Fourier-Hankel spectral method.
For a full review on local helioseismology, see \cite{Gizon2005}.
However, the physical interpretation of these techniques is problematic in
the presence of strong magnetic fields.
The variation and suppression of acoustic power in sunspots causes
anisotropies in the properties of the wave propagation, and therefore
uncertainties in the subsurface flows.
Although time-distance helioseismology and ring-diagram analysis
produce qualitatively similar results \citep{Kosovichev2011}, and although some
uncertainties can be addressed using different algorithms and approximations,
the effects of magnetized subsurface turbulence in the methods is still not well
accounted for, so the separation of magnetic and thermal effects remains open
\citep{Kosovichev2012}.
There, most of the approximations are invalid and we still
lack a reliable technique proven to reproduce the impact of magnetic
fields on the internal traveling waves.
Different studies of local helioseismology measure the flows
around and beneath sunspots, but their results are contradictory
\citep{Jensen2001,Komm2008,Zharkov2008,ZKS2010}.
Studies combining photospheric measurements of the
observed magnetic field vectors and subphotospheric flows derived from
time-distance helioseismology yield similar velocity patters: inward flows in the
sunspot umbra, and outward flows in the surrounding areas.
There are also some differences: the flux emergence-related surface flows, like the
separate motion of leading and following polarity, the fast rotation, or the
apparent shear, do not have their counterpart in the subphotospheric layer.
Other studies try to find subsurface signatures of the emergence of
sunspots before their appearance,
for example \cite{Kosovichev2016} detected
signatures of flux emergence at a depth of 62--75 Mm, 12 hours before the first
bipolar magnetic structures were seen in the photosphere.
Using the same active region, however, different analysis techniques
still give different results \citep{Ilonidis2011,Braun2012}.
Moreover, the attempt to reproduce simulated MHD data from time-distance
helioseismology demonstrates the difficulty to recover horizontal flows near
active regions, and even the inability to recover vertical flows at any
part of the simulation \citep{DeGrave2014}.

Another possible approach in trying to study active regions
and their emergence characteristics is the analysis of
the region just before the emergence of the strong magnetic field,
and the statistical analysis of several such emergences.
\cite{Leka13} compiled a list of pre-emerging active regions and analyzed
them using helioseismic holography \citep{BBLBJ2013, Barnes2014}.
They selected around 100 active regions using GONG and SOHO/MDI, and
computed averages and statistical emergence properties, such as flows.
Statistically,
they could not find large flows of more than $15\m\s^{-1}$ in the top 20 Mm
below the photosphere on the day prior to the visible emergence of the
active region, but could find some signatures of a downflow right before the
emergence;
see also the recent studies by \cite{BSBCGLR16} and \cite{Schunker2016}.

An independent approach is to measure strengthening of the $f$-mode
prior to active region formation \citep{Singh14b}.
It turned out that isolated active regions show a strengthening
some 1--2 days prior to magnetic flux emergence \citep{Singh16}.
Such a slow build-up of subsurface magnetic flux is incompatible
with a buoyant rise and suggestive of an in situ process
such as NEMPI.

Yet another idea is to infer the flows in sunspots and active regions
by studying their surroundings.
This is possible with the Fourier-Hankel
spectral method, which allows the separation of waves into inward
and outward traveling ones, while using only the region around sunspots.
This method was first proposed by \cite{Braun1987}, and applied to
different active regions, pores and the quiet Sun by a number of authors
\citep{Braun1988, Braun1992, Bogdan1993, Braun1995, Crouch2005,Couvidat2013},
and also to the computation of the meridional circulation \citep{Braun1998, Doerr2010}.
These authors agree on the absorption of the incoming waves in the sunspots regions,
which can reach up to 50\% of the $p-$modes power and depends on the depth, frequency,
and ridge mode.

We aim to combine these two approaches, by studying pre-emergence active regions using
Hankel analysis.
Although a comparison between regions with sunspots, pores and quiet-Sun
has been made in these different studies, no comprehensive study
comprises the pre-emergence phase of an active region, and there is a
lack of results for high spherical degree $l$.
Therefore, we hope to complete the picture of Hankel analysis on
sunspots by selecting a group of active region and studying their
emergence properties.
This will be subject of a future publication.

\section{Conclusions}

Even though sunspots and active regions are of vital importance in
solar physics and play an important role in many solar phenomena such
as for example surface magnetic fields, coronal mass ejections, and
solar wind, we lack
a complete theoretical understanding to explain their formation and
evolution.
Does the deeply-rooted flux tube model
provide a complete description of sunspot formation?
Can NEMPI give a possible explanation for a shallow formation process?
Do we have to develop a mixed model where flux tubes emerge and
concentrate via NEMPI? We still lack the means to figure out if
NEMPI is working in MHD simulations like those of \cite{rempel2014}.
Or do other mechanisms play an important role or even explain totally the
formation of active regions?
There are a number of MHD simulations where magnetic structures of different
sizes form spontaneously,
like the thermo-hydromagnetic instability proposed by \cite{Kitchatinov2000},
or the small-scale vortices at intersections of the intergranular lanes can
develop strong vortical downdrafts leading to a spontaneous magnetic flux
concentration \citep{KKWM2010}, or the pore formation in the \cite{SN2012b}
simulations.

NEMPI may well be responsible for concentrating magnetic fields in the shallow layers of
a solar-type simulation.
In view of application to the Sun as a next step, it will be important to consider
more realistic modeling and include the effects of turbulent convection.
It was demonstrated that the relevant
NEMPI parameter $\qp$ is indeed much larger than unity \cite{Kemel12b},
favoring the possibility of NEMPI.
Turbulent convection simulations with an imposed magnetic field
\citep{KBKKR16} yielded structures that are strongly reminiscent of
those found in realistic solar surface simulations in the presence of
full radiative transport \citep{SN2012}.
However, in many existing convection simulations, \citep[see
e.g.][]{KKWM2010,KBKKR16,MS16},
unlike the case of forced turbulence, the scale separation
between the integral scale of the turbulence and the size of the domain
is not large enough for the excitation of NEMPI
and the formation of sharp magnetic structures.
Therefore, the direct detection of negative effective magnetic pressure in
turbulent convection with dynamo-generated magnetic fields
is a difficult problem.
Thus, additional work is needed in order to obtain a complete picture.

Apart from theoretical developments, new and longer high resolution
observations are needed to answer some of these questions.
In particular, it is crucial to have subsurface maps of downflows/upflows before and during
the emergence of the sunspots, or a better description of their structure
deeper down, and an improved treatment of the waves around
strong magnetic fields.

\begin{acknowledgements}

We thank Emanuel Gafton and Ariane Schad for useful discussions and comments on the manuscript.
The research leading to these
results has received funding from the European Research Council
under the European Union’s Seventh Framework Program
(FP/2007-2013)/ERC Grant Agreement no. 307117.
This work was partly supported by
the European Research Council under the
AstroDyn Research Project No.\ 227952,
by the Swedish Research Council under the project grants
621-2011-5076 and 2012-5797 (IRL, AB),
and the Research Council of Norway under the
FRINATEK grant No.\ 231444 (AB, NK, IR).
J.W. acknowledges funding by the Max-Planck/Princeton Center for
Plasma Physics and funding from the People Programme (Marie Curie
Actions) of the European Union's Seventh Framework Programme
(FP7/2007-2013) under REA grant agreement No.\ 623609.
We acknowledge the allocation of computing resources provided by the
Swedish National Allocations Committee at the Center for
Parallel Computers at the Royal Institute of Technology in
Stockholm and the National Supercomputer Centers in Link\"oping and the High
Performance Computing Center North in Ume\aa.
This interdisciplinary collaboration would not have been possible without the support of the Solarnet ``Mobility of Young Researches'' program,
awarded to I.R. Losada.
\end{acknowledgements}

\bibliography{ref}
\bibliographystyle{apalike}

\end{document}

%% file: macros.tex
\newcommand{\EQ}{\begin{equation}}
\newcommand{\EN}{\end{equation}}
\newcommand{\EQA}{\begin{eqnarray}}
\newcommand{\ENA}{\end{eqnarray}}

\newcommand{\Eq}[1]{Equation~(\ref{#1})}

\newcommand{\Fig}[1]{Figure~\ref{#1}}

\newcommand{\Tab}[1]{Table~\ref{#1}}

\newcommand{\meanrho}{\overline{\rho}}

{}
{}
\newcommand{\meanFFFF}{\overline{\mbox{\boldmath ${\cal F}$}}{}}{}

{}
{}
{}
{}
{}
{}
{}
{}
{}
\newcommand{\meanBB}{\overline{\mbox{\boldmath $B$}}{}}{}
{}
\newcommand{\meanFF}{\overline{\mbox{\boldmath $F$}}{}}{}
{}
{}
{}
{}
{}
{}
\newcommand{\meanJJ}{\overline{\mbox{\boldmath $J$}}{}}{}
{}
\newcommand{\meanUU}{\overline{\mbox{\boldmath $U$}}{}}{}

\newcommand{\meanWW}{\overline{\mbox{\boldmath $W$}}{}}{}
{}
{}

\newcommand{\meanp}{\overline{p}}
\newcommand{\meanP}{\overline{P}}

{}

{}
{}

\newcommand{\ww}{\mbox{\boldmath $w$} {}}

\newcommand{\bb}{\mbox{\boldmath $b$} {}}

\newcommand{\uu}{\mbox{\boldmath $u$} {}}
\newcommand{\UU}{\mbox{\boldmath $U$} {}}

\newcommand{\BB}{\mbox{\boldmath $B$} {}}

\newcommand{\JJ}{\mbox{\boldmath $J$} {}}

\newcommand{\AAA}{\mbox{\boldmath $A$} {}}

\newcommand{\ff}{\mbox{\boldmath $f$} {}}

\newcommand{\FF}{\mbox{\boldmath $F$} {}}

\newcommand{\WW}{\mbox{\boldmath $W$} {}}

\newcommand{\grav}{\mbox{\boldmath $g$} {}}
\newcommand{\nab}{\mbox{\boldmath $\nabla$} {}}
\newcommand{\OO}{\mbox{\boldmath $\Omega$} {}}

\newcommand{\SSSS}{\mbox{\boldmath ${\sf S}$} {}}

\newcommand{\DD}{{\rm D} {}}

\def\Co{\mbox{\rm Co}}

\def\Pm{\mbox{\rm Pr}_M}
\def\Rm{\mbox{\rm Re}_M}

\def\Rey{\mbox{\rm Re}}

\def\Co{\mbox{\rm Co}}

\def\cs{c_{\rm s}}

\def\qpz{q_{\rm p0}}
\def\qp{q_{\rm p}}

\def\betap{\beta_{\rm p}}

\def\betastar{\beta_{\star}}

\def\Peff{{\cal P}_{\rm eff}}

\def\kf{k_{\rm f}}

\def\urms{u_{\rm rms}}

\def\nut{\nu_{\rm t}}

\def\etat{\eta_{\rm t}}

\def\Beq{B_{\rm eq}}
\def\Beqz{B_{\rm eq0}}

\def\half{{\textstyle{1\over2}}}

\def\onethird{{\textstyle{1\over3}}}

\newcommand{\G}{\,{\rm G}}

\newcommand{\s}{\,{\rm s}}

\newcommand{\m}{\,{\rm m}}

\newcommand{\Mm}{\,{\rm Mm}}

%% file: aas_macros.tex
\let\jnl@style=\rm
\def\ref@jnl#1{{\jnl@style#1}}

\def\aj{\ref@jnl{AJ}}                   
\def\actaa{\ref@jnl{Acta Astron.}}      
\def\araa{\ref@jnl{ARA\&A}}             
\def\apj{\ref@jnl{ApJ}}                 
\def\apjl{\ref@jnl{ApJL}}                
\def\apjs{\ref@jnl{ApJS}}               
\def\ao{\ref@jnl{Appl.~Opt.}}           
\def\apss{\ref@jnl{Ap\&SS}}             
\def\aap{\ref@jnl{A\&A}}                
\def\aapr{\ref@jnl{A\&A~Rev.}}          
\def\aaps{\ref@jnl{A\&AS}}              
\def\azh{\ref@jnl{AZh}}                 
\def\baas{\ref@jnl{BAAS}}               
\def\bac{\ref@jnl{Bull. astr. Inst. Czechosl.}}
\def\caa{\ref@jnl{Chinese Astron. Astrophys.}}
\def\cjaa{\ref@jnl{Chinese J. Astron. Astrophys.}}
\def\icarus{\ref@jnl{Icarus}}           
\def\jcap{\ref@jnl{J. Cosmology Astropart. Phys.}}
\def\jrasc{\ref@jnl{JRASC}}             
\def\memras{\ref@jnl{MmRAS}}            
\def\mnras{\ref@jnl{MNRAS}}             
\def\na{\ref@jnl{New A}}                
\def\nar{\ref@jnl{New A Rev.}}          
\def\pra{\ref@jnl{Phys.~Rev.~A}}        
\def\prb{\ref@jnl{Phys.~Rev.~B}}        
\def\prc{\ref@jnl{Phys.~Rev.~C}}        
\def\prd{\ref@jnl{Phys.~Rev.~D}}        
\def\pre{\ref@jnl{Phys.~Rev.~E}}        
\def\prl{\ref@jnl{Phys.~Rev.~Lett.}}    
\def\pasa{\ref@jnl{PASA}}               
\def\pasp{\ref@jnl{PASP}}               
\def\pasj{\ref@jnl{PASJ}}               
\def\rmxaa{\ref@jnl{Rev. Mexicana Astron. Astrofis.}}%
\def\qjras{\ref@jnl{QJRAS}}             
\def\skytel{\ref@jnl{S\&T}}             
\def\solphys{\ref@jnl{Sol.~Phys.}}      
\def\sovast{\ref@jnl{Soviet~Ast.}}      
\def\ssr{\ref@jnl{Space~Sci.~Rev.}}     
\def\zap{\ref@jnl{ZAp}}                 
\def\nat{\ref@jnl{Nature}}              
\def\iaucirc{\ref@jnl{IAU~Circ.}}       
\def\aplett{\ref@jnl{Astrophys.~Lett.}} 
\def\apspr{\ref@jnl{Astrophys.~Space~Phys.~Res.}}
\def\bain{\ref@jnl{Bull.~Astron.~Inst.~Netherlands}} 
\def\fcp{\ref@jnl{Fund.~Cosmic~Phys.}}  
\def\gca{\ref@jnl{Geochim.~Cosmochim.~Acta}}   
\def\grl{\ref@jnl{Geophys.~Res.~Lett.}} 
\def\jcp{\ref@jnl{J.~Chem.~Phys.}}      
\def\jgr{\ref@jnl{J.~Geophys.~Res.}}    
\def\jqsrt{\ref@jnl{J.~Quant.~Spec.~Radiat.~Transf.}}
\def\memsai{\ref@jnl{Mem.~Soc.~Astron.~Italiana}}
\def\nphysa{\ref@jnl{Nucl.~Phys.~A}}   
\def\physrep{\ref@jnl{Phys.~Rep.}}   
\def\physscr{\ref@jnl{Phys.~Scr}}   
\def\planss{\ref@jnl{Planet.~Space~Sci.}}   
\def\procspie{\ref@jnl{Proc.~SPIE}}   

%% file: IAU237_IRL.bbl
\begin{thebibliography}{}

\bibitem[{Babcock}, 1961]{Babcock61}
{Babcock}, H.~W. (1961).
\newblock {The Topology of the Sun's Magnetic Field and the 22-Year Cycle.}
\newblock {\em \apj}, 133:572.

\bibitem[{Barnes} et~al., 2014]{Barnes2014}
{Barnes}, G., {Birch}, A.~C., {Leka}, K.~D., and {Braun}, D.~C. (2014).
\newblock {Helioseismology of Pre-emerging Active Regions. III. Statistical
  Analysis}.
\newblock {\em \apj}, 786:19.

\bibitem[{Birch} et~al., 2013]{BBLBJ2013}
{Birch}, A.~C., {Braun}, D.~C., {Leka}, K.~D., {Barnes}, G., and {Javornik}, B.
  (2013).
\newblock {Helioseismology of Pre-emerging Active Regions. II. Average
  Emergence Properties}.
\newblock {\em \apj}, 762:131.

\bibitem[Birch et~al., 2016]{BSBCGLR16}
Birch, A.~C., Schunker, H., Braun, D.~C., Cameron, R., Gizon, L., L{\"o}ptien,
  B., and Rempel, M. (2016).
\newblock A low upper limit on the subsurface rise speed of solar active
  regions.
\newblock {\em Science Advances}, 2(7).

\bibitem[{Bogdan} et~al., 1993]{Bogdan1993}
{Bogdan}, T.~J., {Brown}, T.~M., {Lites}, B.~W., and {Thomas}, J.~H. (1993).
\newblock {The absorption of p-modes by sunspots - Variations with degree and
  order}.
\newblock {\em \apj}, 406:723--734.

\bibitem[{Brandenburg} et~al., 2011]{BKKMR11}
{Brandenburg}, A., {Kemel}, K., {Kleeorin}, N., {Mitra}, D., and
  {Rogachevskii}, I. (2011).
\newblock {Detection of Negative Effective Magnetic Pressure Instability in
  Turbulence Simulations}.
\newblock {\em \apjl}, 740:L50.

\bibitem[{Brandenburg} et~al., 2012]{BKKR12}
{Brandenburg}, A., {Kemel}, K., {Kleeorin}, N., and {Rogachevskii}, I. (2012).
\newblock {The Negative Effective Magnetic Pressure in Stratified Forced
  Turbulence}.
\newblock {\em \apj}, 749:179.

\bibitem[{Brandenburg} et~al., 2013]{BKR13}
{Brandenburg}, A., {Kleeorin}, N., and {Rogachevskii}, I. (2013).
\newblock {Self-assembly of Shallow Magnetic Spots through Strongly Stratified
  Turbulence}.
\newblock {\em \apjl}, 776:L23.

\bibitem[{Brandenburg} et~al., 2016]{BRK16}
{Brandenburg}, A., {Rogachevskii}, I., and {Kleeorin}, N. (2016).
\newblock {Magnetic concentrations in stratified turbulence: the negative
  effective magnetic pressure instability}.
\newblock {\em New Journal of Physics}, 18(12):125011.

\bibitem[{Brandenburg} and {Subramanian}, 2005]{BS05}
{Brandenburg}, A. and {Subramanian}, K. (2005).
\newblock {Astrophysical magnetic fields and nonlinear dynamo theory}.
\newblock {\em \physrep}, 417:1--209.

\bibitem[{Braun}, 1995]{Braun1995}
{Braun}, D.~C. (1995).
\newblock {Scattering of p-Modes by Sunspots. I. Observations}.
\newblock {\em \apj}, 451:859.

\bibitem[{Braun}, 2012]{Braun2012}
{Braun}, D.~C. (2012).
\newblock {Comment on "Detection of Emerging Sunspot Regions in the Solar
  Interior"}.
\newblock {\em Science}, 336:296.

\bibitem[{Braun} et~al., 1987]{Braun1987}
{Braun}, D.~C., {Duvall}, Jr., T.~L., and {Labonte}, B.~J. (1987).
\newblock {Acoustic absorption by sunspots}.
\newblock {\em \apjl}, 319:L27--L31.

\bibitem[{Braun} et~al., 1988]{Braun1988}
{Braun}, D.~C., {Duvall}, Jr., T.~L., and {Labonte}, B.~J. (1988).
\newblock {The absorption of high-degree p-mode oscillations in and around
  sunspots}.
\newblock {\em \apj}, 335:1015--1025.

\bibitem[{Braun} et~al., 1992]{Braun1992}
{Braun}, D.~C., {Duvall}, Jr., T.~L., {Labonte}, B.~J., {Jefferies}, S.~M.,
  {Harvey}, J.~W., and {Pomerantz}, M.~A. (1992).
\newblock {Scattering of p-modes by a sunspot}.
\newblock {\em \apjl}, 391:L113--L116.

\bibitem[{Braun} and {Fan}, 1998]{Braun1998}
{Braun}, D.~C. and {Fan}, Y. (1998).
\newblock {Helioseismic Measurements of the Subsurface Meridional Flow}.
\newblock {\em \apjl}, 508:L105--L108.

\bibitem[{Caligari} et~al., 1995]{caligari95}
{Caligari}, P., {Moreno-Insertis}, F., and {Schussler}, M. (1995).
\newblock {Emerging flux tubes in the solar convection zone. 1: Asymmetry,
  tilt, and emergence latitude}.
\newblock {\em \apj}, 441:886--902.

\bibitem[{Cheung} and {Isobe}, 2014]{Cheung+Isobe:14}
{Cheung}, M.~C.~M. and {Isobe}, H. (2014).
\newblock {Flux Emergence (Theory)}.
\newblock {\em Living Reviews in Solar Physics}, 11:3.

\bibitem[{Clette} et~al., 2014]{Clette14}
{Clette}, F., {Svalgaard}, L., {Vaquero}, J.~M., and {Cliver}, E.~W. (2014).
\newblock {Revisiting the Sunspot Number. A 400-Year Perspective on the Solar
  Cycle}.
\newblock {\em \ssr}, 186:35--103.

\bibitem[{Couvidat}, 2013]{Couvidat2013}
{Couvidat}, S. (2013).
\newblock {Oscillation Power in Sunspots and Quiet Sun from Hankel Analysis
  Performed on SDO/HMI and SDO/AIA Data}.
\newblock {\em \solphys}, 282:15--38.

\bibitem[{Crouch} et~al., 2005]{Crouch2005}
{Crouch}, A.~D., {Cally}, P.~S., {Charbonneau}, P., {Braun}, D.~C., and
  {Desjardins}, M. (2005).
\newblock {Genetic magnetohelioseismology with Hankel analysis data}.
\newblock {\em \mnras}, 363:1188--1204.

\bibitem[{DeGrave} et~al., 2014]{DeGrave2014}
{DeGrave}, K., {Jackiewicz}, J., and {Rempel}, M. (2014).
\newblock {Time-distance Helioseismology of Two Realistic Sunspot Simulations}.
\newblock {\em \apj}, 794:18.

\bibitem[{Dikpati} and {Gilman}, 2001]{Dikpati2001}
{Dikpati}, M. and {Gilman}, P.~A. (2001).
\newblock {Flux-Transport Dynamos with {$\alpha$}-Effect from Global
  Instability of Tachocline Differential Rotation: A Solution for Magnetic
  Parity Selection in the Sun}.
\newblock {\em \apj}, 559:428--442.

\bibitem[{Doerr} et~al., 2010]{Doerr2010}
{Doerr}, H.-P., {Roth}, M., {Zaatri}, A., {Krieger}, L., and {Thompson}, M.~J.
  (2010).
\newblock {A new code for Fourier-Legendre analysis of large datasets: First
  results and a comparison with ring-diagram analysis}.
\newblock {\em Astron. Nachr.}, 331:911.

\bibitem[{Gizon} and {Birch}, 2005]{Gizon2005}
{Gizon}, L. and {Birch}, A.~C. (2005).
\newblock {Local Helioseismology}.
\newblock {\em Living Reviews in Solar Physics}, 2:6.

\bibitem[{Ilonidis} et~al., 2011]{Ilonidis2011}
{Ilonidis}, S., {Zhao}, J., and {Kosovichev}, A. (2011).
\newblock {Detection of Emerging Sunspot Regions in the Solar Interior}.
\newblock {\em Science}, 333:993.

\bibitem[{Jabbari} et~al., 2014]{Jabbari2014}
{Jabbari}, S., {Brandenburg}, A., {Losada}, I.~R., {Kleeorin}, N., and
  {Rogachevskii}, I. (2014).
\newblock {Magnetic flux concentrations from dynamo-generated fields}.
\newblock {\em \aap}, 568:A112.

\bibitem[{Jensen} et~al., 2001]{Jensen2001}
{Jensen}, J.~M., {Duvall}, Jr., T.~L., {Jacobsen}, B.~H., and
  {Christensen-Dalsgaard}, J. (2001).
\newblock {Imaging an Emerging Active Region with Helioseismic Tomography}.
\newblock {\em \apjl}, 553:L193--L196.

\bibitem[{K{\"a}pyl{\"a}} et~al., 2016]{KBKKR16}
{K{\"a}pyl{\"a}}, P.~J., {Brandenburg}, A., {Kleeorin}, N., {K{\"a}pyl{\"a}},
  M.~J., and {Rogachevskii}, I. (2016).
\newblock {Magnetic flux concentrations from turbulent stratified convection}.
\newblock {\em \aap}, 588:A150.

\bibitem[{Kemel} et~al., 2012a]{Kemel12b}
{Kemel}, K., {Brandenburg}, A., {Kleeorin}, N., {Mitra}, D., and
  {Rogachevskii}, I. (2012a).
\newblock {Spontaneous Formation of Magnetic Flux Concentrations in Stratified
  Turbulence}.
\newblock {\em \solphys}, 280:321--333.

\bibitem[{Kemel} et~al., 2013]{Kemel13a}
{Kemel}, K., {Brandenburg}, A., {Kleeorin}, N., {Mitra}, D., and
  {Rogachevskii}, I. (2013).
\newblock {Active Region Formation through the Negative Effective Magnetic
  Pressure Instability}.
\newblock {\em \solphys}, 287:293--313.

\bibitem[{Kemel} et~al., 2012b]{Kemel12a}
{Kemel}, K., {Brandenburg}, A., {Kleeorin}, N., and {Rogachevskii}, I. (2012b).
\newblock {Properties of the negative effective magnetic pressure instability}.
\newblock {\em Astron. Nachr.}, 333:95.

\bibitem[{Kitchatinov} and {Mazur}, 2000]{Kitchatinov2000}
{Kitchatinov}, L.~L. and {Mazur}, M.~V. (2000).
\newblock {Stability and equilibrium of emerged magnetic flux}.
\newblock {\em \solphys}, 191:325--340.

\bibitem[{Kitiashvili} et~al., 2010]{KKWM2010}
{Kitiashvili}, I.~N., {Kosovichev}, A.~G., {Wray}, A.~A., and {Mansour}, N.~N.
  (2010).
\newblock {Mechanism of Spontaneous Formation of Stable Magnetic Structures on
  the Sun}.
\newblock {\em \apj}, 719:307--312.

\bibitem[{Kleeorin} et~al., 1993]{KMR93}
{Kleeorin}, N., {Mond}, M., and {Rogachevskii}, I. (1993).
\newblock {Magnetohydrodynamic instabilities in developed small-scale
  turbulence}.
\newblock {\em Physics of Fluids B}, 5:4128--4134.

\bibitem[{Kleeorin} et~al., 1996]{KMR96}
{Kleeorin}, N., {Mond}, M., and {Rogachevskii}, I. (1996).
\newblock {Magnetohydrodynamic turbulence in the solar convective zone as a
  source of oscillations and sunspots formation.}
\newblock {\em \aap}, 307:293.

\bibitem[{Kleeorin} and {Rogachevskii}, 1994]{KR94}
{Kleeorin}, N. and {Rogachevskii}, I. (1994).
\newblock {Effective Amp{\`e}re force in developed magnetohydrodynamic
  turbulence}.
\newblock {\em \pre}, 50:2716--2730.

\bibitem[{Kleeorin} et~al., 1990]{KRR90}
{Kleeorin}, N., {Rogachevskii}, I., and {Ruzmaikin}, A. (1990).
\newblock {Magnetic force reversal and instability in a plasma with developed
  magnetohydrodynamic turbulence}.
\newblock {\em JETP}, 70:878--883.

\bibitem[{Kleeorin} et~al., 1989]{KRR89}
{Kleeorin}, N.~I., {Rogachevskii}, I.~V., and {Ruzmaikin}, A.~A. (1989).
\newblock {The effect of negative magnetic pressure and the large-scale
  magnetic field instability in the solar convective zone}.
\newblock {\em Pisma v Astronomicheskii Zhurnal}, 15:639--645.

\bibitem[{Komm} et~al., 2008]{Komm2008}
{Komm}, R., {Morita}, S., {Howe}, R., and {Hill}, F. (2008).
\newblock {Emerging Active Regions Studied with Ring-Diagram Analysis}.
\newblock {\em \apj}, 672:1254--1265.

\bibitem[{Kosovichev}, 2012]{Kosovichev2012}
{Kosovichev}, A.~G. (2012).
\newblock {Local Helioseismology of Sunspots: Current Status and Perspectives}.
\newblock {\em \solphys}, 279:323--348.

\bibitem[{Kosovichev} et~al., 2011]{Kosovichev2011}
{Kosovichev}, A.~G., {Basu}, S., {Bogart}, R., {Duvall}, Jr., T.~L.,
  {Gonzalez-Hernandez}, I., {Haber}, D., {Hartlep}, T., {Howe}, R., {Komm}, R.,
  {Kholikov}, S., {Parchevsky}, K.~V., {Tripathy}, S., and {Zhao}, J. (2011).
\newblock {Local helioseismology of sunspot regions: Comparison of ring-diagram
  and time-distance results}.
\newblock In {\em Journal of Physics Conference Series}, volume 271 of {\em
  Journal of Physics Conference Series}, page 012005.

\bibitem[{Kosovichev} et~al., 2016]{Kosovichev2016}
{Kosovichev}, A.~G., {Zhao}, J., and {Ilonidis}, S. (2016).
\newblock {Local Helioseismology of Emerging Active Regions: A Case Study}.
\newblock {\em ArXiv e-prints}.

\bibitem[{Krause} and {R\"adler}, 1980]{KR80}
{Krause}, F. and {R\"adler}, K.-H. (1980).
\newblock {\em {Mean-field magnetohydrodynamics and dynamo theory}}.

\bibitem[{Leighton}, 1964]{Leighton64}
{Leighton}, R.~B. (1964).
\newblock {Transport of Magnetic Fields on the Sun.}
\newblock {\em \apj}, 140:1547.

\bibitem[{Leighton}, 1969]{Leighton69}
{Leighton}, R.~B. (1969).
\newblock {A Magneto-Kinematic Model of the Solar Cycle}.
\newblock {\em \apj}, 156:1.

\bibitem[{Leka} et~al., 2013]{Leka13}
{Leka}, K.~D., {Barnes}, G., {Birch}, A.~C., {Gonzalez-Hernandez}, I., {Dunn},
  T., {Javornik}, B., and {Braun}, D.~C. (2013).
\newblock {Helioseismology of Pre-emerging Active Regions. I. Overview, Data,
  and Target Selection Criteria}.
\newblock {\em \apj}, 762:130.

\bibitem[{Losada} et~al., 2012]{LBKMR2012}
{Losada}, I.~R., {Brandenburg}, A., {Kleeorin}, N., {Mitra}, D., and
  {Rogachevskii}, I. (2012).
\newblock {Rotational effects on the negative magnetic pressure instability}.
\newblock {\em \aap}, 548:A49.

\bibitem[{Losada} et~al., 2013]{LBKR2013a}
{Losada}, I.~R., {Brandenburg}, A., {Kleeorin}, N., and {Rogachevskii}, I.
  (2013).
\newblock {Competition of rotation and stratification in flux concentrations}.
\newblock {\em \aap}, 556:A83.

\bibitem[{Masada} and {Sano}, 2016]{MS16}
{Masada}, Y. and {Sano}, T. (2016).
\newblock {Spontaneous Formation of Surface Magnetic Structure from Large-scale
  Dynamo in Strongly Stratified Convection}.
\newblock {\em \apjl}, 822:L22.

\bibitem[{Moffatt}, 1978]{Mof78}
{Moffatt}, H.~K. (1978).
\newblock {\em {Magnetic Field Generation in Electrically Conducting Fluids}}.
\newblock Cambridge University Press, Cambridge.

\bibitem[{Parker}, 1955]{Par55b}
{Parker}, E.~N. (1955).
\newblock {The Formation of Sunspots from the Solar Toroidal Field.}
\newblock {\em \apj}, 121:491.

\bibitem[{Parker}, 1975]{Parker75}
{Parker}, E.~N. (1975).
\newblock {The generation of magnetic fields in astrophysical bodies. X -
  Magnetic buoyancy and the solar dynamo}.
\newblock {\em \apj}, 198:205--209.

\bibitem[{Rempel} and {Cheung}, 2014]{rempel2014}
{Rempel}, M. and {Cheung}, M.~C.~M. (2014).
\newblock {Numerical Simulations of Active Region Scale Flux Emergence: From
  Spot Formation to Decay}.
\newblock {\em \apj}, 785:90.

\bibitem[{Rogachevskii} and {Kleeorin}, 2007]{RK07}
{Rogachevskii}, I. and {Kleeorin}, N. (2007).
\newblock {Magnetic fluctuations and formation of large-scale inhomogeneous
  magnetic structures in a turbulent convection}.
\newblock {\em \pre}, 76(5):056307.

\bibitem[{Schunker} et~al., 2016]{Schunker2016}
{Schunker}, H., {Braun}, D.~C., {Birch}, A.~C., {Burston}, R.~B., and {Gizon},
  L. (2016).
\newblock {SDO/HMI survey of emerging active regions for helioseismology}.
\newblock {\em \aap}, 595:A107.

\bibitem[{Singh} et~al., 2014]{Singh14b}
{Singh}, N.~K., {Brandenburg}, A., and {Rheinhardt}, M. (2014).
\newblock {Fanning Out of the Solar f-mode in the Presence of Non-uniform
  Magnetic Fields?}
\newblock {\em \apjl}, 795:L8.

\bibitem[{Singh} et~al., 2016]{Singh16}
{Singh}, N.~K., {Raichur}, H., and {Brandenburg}, A. (2016).
\newblock {High-wavenumber Solar f-mode Strengthening Prior to Active Region
  Formation}.
\newblock {\em \apj}, 832:120.

\bibitem[{Spruit}, 1974]{Spruit74}
{Spruit}, H.~C. (1974).
\newblock {A model of the solar convection zone}.
\newblock {\em \solphys}, 34:277--290.

\bibitem[{Steenbeck} et~al., 1966]{SKR66}
{Steenbeck}, M., {Krause}, F., and {R{\"a}dler}, K.-H. (1966).
\newblock {Berechnung der mittleren Lorentz-Feldst{\"a}rke $\overline{\vv
  \times \bb}$ f{\"u}r ein elektrisch leitendes Medium in turbulenter, durch
  Coriolis-Kr{\"a}fte beeinflu{\ss}ter Bewegung}.
\newblock {\em Z. Naturforsch. A}, 21:369.

\bibitem[{Stein} and {Nordlund}, 2012a]{SN2012b}
{Stein}, R. and {Nordlund}, A. (2012a).
\newblock {Spontaneous Pore Formation in Magneto-Convection Simulations}.
\newblock In {Golub}, L., {De Moortel}, I., and {Shimizu}, T., editors, {\em
  Fifth Hinode Science Meeting}, volume 456 of {\em Astronomical Society of the
  Pacific Conference Series}, page~39.

\bibitem[{Stein} and {Nordlund}, 2012b]{SN2012}
{Stein}, R.~F. and {Nordlund}, {\AA}. (2012b).
\newblock {On the Formation of Active Regions}.
\newblock {\em \apjl}, 753:L13.

\bibitem[{Stix}, 1976]{Stix1976}
{Stix}, M. (1976).
\newblock {Differential rotation and the solar dynamo}.
\newblock {\em \aap}, 47:243--254.

\bibitem[{Warnecke} and {Brandenburg}, 2010]{WB2010}
{Warnecke}, J. and {Brandenburg}, A. (2010).
\newblock {Surface appearance of dynamo-generated large-scale fields}.
\newblock {\em \aap}, 523:A19.

\bibitem[{Warnecke} et~al., 2011]{WBM2011}
{Warnecke}, J., {Brandenburg}, A., and {Mitra}, D. (2011).
\newblock {Dynamo-driven plasmoid ejections above a spherical surface}.
\newblock {\em \aap}, 534:A11.

\bibitem[{Warnecke} et~al., 2016a]{WKKB16}
{Warnecke}, J., {K{\"a}pyl{\"a}}, P.~J., {K{\"a}pyl{\"a}}, M.~J., and
  {Brandenburg}, A. (2016a).
\newblock {Influence of a coronal envelope as a free boundary to global
  convective dynamo simulations}.
\newblock {\em \aap}, 596:A115.

\bibitem[{Warnecke} et~al., 2012]{WKMB12}
{Warnecke}, J., {K{\"a}pyl{\"a}}, P.~J., {Mantere}, M.~J., and {Brandenburg},
  A. (2012).
\newblock {Ejections of Magnetic Structures Above a Spherical Wedge Driven by a
  Convective Dynamo with Differential Rotation}.
\newblock {\em \solphys}, 280:299--319.

\bibitem[{Warnecke} et~al., 2013a]{WKMB13}
{Warnecke}, J., {K{\"a}pyl{\"a}}, P.~J., {Mantere}, M.~J., and {Brandenburg},
  A. (2013a).
\newblock {Spoke-like Differential Rotation in a Convective Dynamo with a
  Coronal Envelope}.
\newblock {\em \apj}, 778:141.

\bibitem[{Warnecke} et~al., 2013b]{WLBKR2013}
{Warnecke}, J., {Losada}, I.~R., {Brandenburg}, A., {Kleeorin}, N., and
  {Rogachevskii}, I. (2013b).
\newblock {Bipolar magnetic structures driven by stratified turbulence with a
  coronal envelope}.
\newblock {\em \apjl}, 777:L37.

\bibitem[{Warnecke} et~al., 2016b]{WLBKR2015}
{Warnecke}, J., {Losada}, I.~R., {Brandenburg}, A., {Kleeorin}, N., and
  {Rogachevskii}, I. (2016b).
\newblock {Bipolar region formation in stratified two-layer turbulence}.
\newblock {\em \aap}, 589:A125.

\bibitem[{Zhao} et~al., 2010]{ZKS2010}
{Zhao}, J., {Kosovichev}, A.~G., and {Sekii}, T. (2010).
\newblock {High-Resolution Helioseismic Imaging of Subsurface Structures and
  Flows of a Solar Active Region Observed by Hinode}.
\newblock {\em \apj}, 708:304--313.

\bibitem[{Zharkov} and {Thompson}, 2008]{Zharkov2008}
{Zharkov}, S. and {Thompson}, M.~J. (2008).
\newblock {Time Distance Analysis of the Emerging Active Region NOAA 10790}.
\newblock {\em \solphys}, 251:369--380.

\end{thebibliography}
